\documentclass[conference]{IEEEtran}
\IEEEoverridecommandlockouts

\usepackage{cmap}
\usepackage[T1]{fontenc}
\usepackage{cite}
\usepackage[ruled,linesnumbered]{algorithm2e}
\usepackage{amsmath,amssymb,amsfonts}
\usepackage{algorithmic}
\usepackage{graphicx}
\usepackage{textcomp}
\usepackage{xcolor}
\usepackage{bm}
\usepackage{amsthm}
\usepackage{enumerate}
\usepackage{booktabs}
\usepackage{diagbox}
\usepackage{supertabular}
\usepackage{subfigure}
\usepackage{caption}

\newtheorem{definition}{Definition}

\def\BibTeX{{\rm B\kern-.05em{\sc i\kern-.025em b}\kern-.08em
    T\kern-.1667em\lower.7ex\hbox{E}\kern-.125emX}}
\begin{document}

\title{Learning-based Big Data Sharing Incentive in
Mobile AIGC Networks}

\author{
\IEEEauthorblockN{Jinbo Wen\IEEEauthorrefmark{1}, Yang Zhang\IEEEauthorrefmark{1}, Yulin Chen\IEEEauthorrefmark{2}, Weifeng Zhong\IEEEauthorrefmark{2}, Xumin Huang\IEEEauthorrefmark{2}, Lei Liu\IEEEauthorrefmark{3}, Dusit Niyato\IEEEauthorrefmark{4}, \textit{Fellow, IEEE}}

\IEEEcompsocitemizethanks{

\textit{Corresponding author: Yang Zhang (e-mail: yangzhang@nuaa.edu.cn).}

}

\IEEEauthorblockA{\IEEEauthorrefmark{1}\textit{College of Computer Science and Technology, Nanjing University of Aeronautics and Astronautics, China}\\
\IEEEauthorrefmark{2}\textit{School of Automation, Guangdong University of Technology, China}\\
\IEEEauthorrefmark{3}\textit{School of Telecommunications Engineering, Xidian University, China}\\
\IEEEauthorrefmark{4}\textit{School of Computer Science and Engineering, Nanyang Technological University, Singapore}\\
}
}

\maketitle

\begin{abstract}
Rapid advancements in wireless communication have led to a dramatic upsurge in data volumes within mobile edge networks. These substantial data volumes offer opportunities for training Artificial Intelligence-Generated Content (AIGC) models to possess strong prediction and decision-making capabilities. AIGC represents an innovative approach that utilizes sophisticated generative AI algorithms to automatically generate diverse content based on user inputs. Leveraging mobile edge networks, mobile AIGC networks enable customized and real-time AIGC services for users by deploying AIGC models on edge devices. Nonetheless, several challenges hinder the provision of high-quality AIGC services, including issues related to the quality of sensing data for AIGC model training and the establishment of incentives for big data sharing from mobile devices to edge devices amidst information asymmetry. In this paper, we initially define a Quality of Data (QoD) metric based on the age of information to quantify the quality of sensing data. Subsequently, we propose a contract theoretic model aimed at motivating mobile devices for big data sharing. Furthermore, we employ a Proximal Policy Optimization (PPO) algorithm to determine the optimal contract. Numerical results demonstrate the efficacy and reliability of the proposed PPO-based contract model.
\end{abstract}

\begin{IEEEkeywords}
Mobile AIGC networks, big data sharing, contract theory, age of information, proximal policy optimization.
\end{IEEEkeywords}


\section{Introduction}
Due to the rapid evolution of cutting-edge technologies like sixth-generation technologies and the Internet of Things, mobile edge networks, characterized by caching and edge computing capabilities at the periphery of cellular networks\cite{wang2017survey}, have witnessed a significant surge in sensing data volumes from large-scale mobile devices\cite{lai2023resource}. The global mobile data traffic market, estimated at 84.1 million TB per month in 2022, is projected to escalate to 603.5 million TB per month by 2030, with mobile data traffic from mobile devices anticipated to reach 257.1 GB per month by the end of 2030, marking a fifty-fold increase from 2010 levels\cite{lai2023resource}. Big data stands as a pivotal modern technology reshaping various application fields\cite{jagatheesaperumal2021duo}. Leveraging big data technologies such as data normalization and feature engineering enables the analysis and processing of large volumes of heterogeneous and unstructured data within mobile edge networks, thus identifying useful patterns for training Artificial Intelligence (AI) models\cite{jagatheesaperumal2021duo}, enabling intelligent mobile services, such as navigation and location-based services, virtual assistants, and AI-Generated Content (AIGC) services\cite{wen2023freshness, xu2024unleashing}.

AIGC utilizes generative AI techniques to autonomously produce content based on user inputs, i.e., prompts, which can complement or replace human content generation\cite{xu2024unleashing, wen2023generative}. Notably, GPT-4 represents the latest milestone for large language models\cite{liu2024sora}, which is a large multimodal model that accepts both image and text inputs to generate original content through human-machine interaction. Sora, among the leading AIGC models, can generate imaginative and realistic videos up to a minute long based on user-provided prompts\cite{liu2024sora}. To mitigate the AIGC service latency inherent in centralized AIGC frameworks\cite{wen2023freshness,liao2024optimizing}, the integration of AIGC and mobile edge networks yields mobile AIGC networks, facilitating real-time and customized content delivery to diminish AIGC service latency and enhance user experiences across diverse applications\cite{xu2024unleashing,wen2023freshness}. The big data within mobile AIGC networks consists of an array of structured and unstructured data captured by mobile devices and smart sources\cite{jagatheesaperumal2021duo}, including text and videos. AIGC models can learn patterns and user preferences from these data and provide customized and real-time AIGC services to users.

Although mobile AIGC networks can provide large amounts of data for AIGC model training and fine-tuning, there are still challenges in achieving high-quality and reliable mobile AIGC services as follows:
\begin{description}
\item[\textbf{C1)}] \textit{How to quantify the quality of sensing data for AIGC model fine-tuning?} The training data quality significantly influences the efficacy of AIGC model learning and the extraction of valuable insights\cite{jagatheesaperumal2021duo}. Biased or outdated information within the training data can lead to inaccuracies during the fine-tuning process, resulting in low-quality AIGC services\cite{wen2023generative}.
\item[\textbf{C2)}] \textit{How to ensure that mobile devices provide high-quality data to edge devices?} Due to the high cost of collecting data in mobile networks\cite{xu2024unleashing}, mobile devices may not be willing to contribute high-quality data without appropriate incentives\cite{wen2023freshness}, or even maliciously provide harmful data to edge devices due to information asymmetry, affecting the quality and reliability of AIGC model training and fine-tuning.
\item[\textbf{C3)}] \textit{How to handle complex decision-making processes within mobile AIGC networks?} Due to the high-dimensional environment of mobile AIGC networks, the decision-making processes in these networks are complex, and traditional mathematical techniques may ineffectively find the optimal policies, such as optimal incentive mechanism design.
\end{description}
Some studies have been conducted to formulate incentive mechanisms for AIGC service provisions\cite{wen2023freshness, lin2024blockchain}. For example, in \cite{wen2023freshness}, the authors utilized contract theory to motivate unmanned aerial vehicles to contribute fresh sensing data for AIGC fine-tuning and inferences. However, most studies do not consider quantifying sensing data quality for AIGC models and adopting AI techniques to find the optimal strategies to adapt to dynamic scenarios.

To tackle the challenges outlined above, we first define a Quality of Data (QoD) metric. Then, we formulate an incentive mechanism to motivate mobile devices to provide fresh data to edge servers and utilize Deep Reinforcement Learning (DRL) to determine the optimal strategy. Our contributions can be summarized as follows:
\begin{itemize}
    \item By jointly considering service latency and Age of Information (AoI) in big data sharing, we propose a QoD metric to quantify the quality of sensing data for AIGC model training, where AoI is widely used to measure data freshness in information systems (for \textbf{C1}).
    \item We propose a contract theoretic model to incentivize mobile devices to provide high-quality sensing data for AIGC models. The optimization objective centers on maximizing the overall utility of the edge server under information asymmetry (for \textbf{C2}).
    \item We adopt Proximal Policy Optimization (PPO) to determine the optimal contract design. Our numerical results show that the proposed PPO-based algorithm has high performance and can effectively adapt to the dynamic states of mobile AIGC networks (for \textbf{C3}).
    
\end{itemize}

\section{System Model}
\begin{figure}[t]
\centerline{\includegraphics[width=0.4\textwidth]{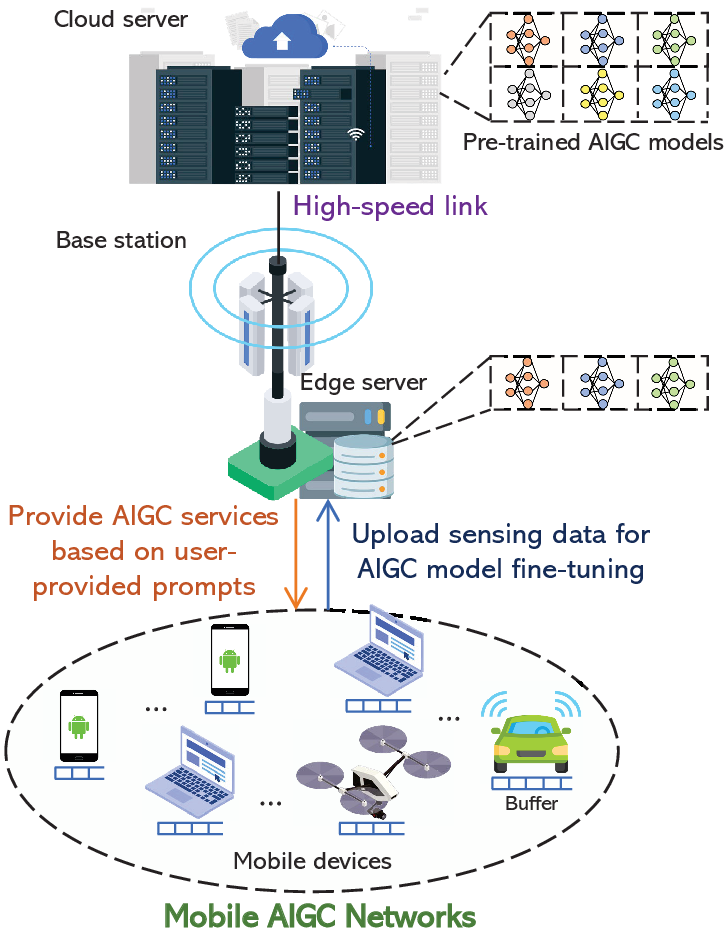}}
\captionsetup{font=footnotesize}
\caption{An illustration of big data sharing in mobile AIGC networks.}
\label{framework}
\end{figure}

In Fig. \ref{framework}, we present an illustration of big data sharing in mobile AIGC networks. In these networks, large-scale mobile devices with sensors can collect large amounts of sensing data from the surrounding environments. For example, the sensors installed in mobile devices can collect data on the speed, location, and even physical conditions of users. These data are cached and periodically updated to ensure the freshness of content\cite{zhang2018towards}. After receiving data requests from Base Stations (BSs), mobile devices can upload these data to edge servers deployed at the BS. To minimize AIGC service delay, edge servers acquire pre-trained AIGC models from cloud data centers. Based on the user requirements, edge servers can fine-tune the pre-trained AIGC models to specific tasks\cite{xu2024unleashing}, where fine-tuning tasks involve processing and analyzing freshly collected datasets from mobile devices to guarantee the quality of AIGC services\cite{wen2023freshness}. Leveraging these trained AIGC models, edge servers perform inferences based on the prompts of users, thereby delivering AIGC services to users. 

However, the private information of mobile devices over big data sharing causes an information asymmetry situation. Given the crucial role of high-quality data in fine-tuning AIGC models for the implementation of superior and customized AIGC models\cite{xu2024unleashing}, BSs can incentivize mobile devices to provide fresh sensing data under information asymmetry. This approach ensures the quality and reliability of AIGC services within mobile AIGC networks.

\section{Contract Modelling}
In this paper, we consider a BS and $M$ UAVs, where $\mathcal{M} = \small\{1,\ldots,m,\ldots,M\small\}$. Initially, we employ the AoI metric to gauge the freshness of sensing data utilized for the fine-tuning of pre-trained AIGC models\cite{wen2023freshness,zhang2018towards}. Subsequently, we propose a contract theoretic model to incentivize mobile devices to furnish high-quality datasets for AIGC model fine-tuning. In our scenario, we consider the utilization of Orthogonal Frequency Division Multiplexing Access (OFDMA) technology in big data sharing, ensuring the orthogonality of all communication channels utilized by various mobile devices and the BS\cite{wen2023task}.

\subsection{Quality of Data Metric}
AoI is used to capture data freshness, which is defined as the duration elapsed since the generation of the most recently received data\cite{liao2024optimizing,zhou2021towards}. The smaller the AoI, the fresher the sensing data, which in turn can be used to train AIGC models for excellent performance. For the quality of sensing data used for AIGC model fine-tuning, the BS not only considers AoI but also service latency in big data sharing. Similar to \cite{zhou2021towards}, we define the QoD by incorporating both service latency and AoI. This metric applies to scenarios where data is periodically updated and cached. In our paper, AoI is defined as the time taken from the collection of sensing data by mobile devices to successful reception at the BS, while service latency is defined as the time taken from the start of a data request to successful reception by the BS.

Defining $l$ as the size of cached data in bytes and $\tau$ as the achievable data transmission rate between the BS and mobile devices\cite{zhang2018towards}, we can obtain $t = l/\tau$, representing the duration of one time slot\cite{zhang2018towards,zhou2021towards}. Mobile device $m$ periodically updates its cached data to maintain the freshness of sensing data, with $\theta_m$ denoting the number of time slots for an update cycle\cite{wen2023freshness}. The cached data is refreshed in the first time slot\cite{zhang2018towards}. As stated in \cite{zhang2018towards}, the AoI for a data request made in the $i$-th time slot is $(i+1)t$, where $i = 2,\ldots,\theta_m-1$, while for a request initiated in the first or last time slot, the AoI is $2t$. Due to the nature of the Poisson process\cite{zhang2018towards,wen2023freshness}, data requests have the same probability of occurring no matter in which time slot, expressed as $1/\theta_m$, the average AoI for data sharing of mobile device $m$ is given by
\begin{equation}\label{AoI}
    \begin{split}
        \overline{A}_m (\theta_m)= \frac{2}{\theta_m}(2t) + \sum_{i = 2}^{\theta_m-1}\frac{(i + 1)t}{\theta_m}= t\bigg(\frac{1}{\theta_m}+\frac{\theta_m}{2}+\frac{1}{2}\bigg).
    \end{split}
\end{equation}

Likewise, when the BS initiates a data request within the first $(\theta_m -1)$ time slots, mobile device $m$ can immediately upload the data to the BS, resulting in a service latency of $t$. Conversely, if the data request occurs at the last time slot, an additional time slot is necessary for data updating, leading to a service latency of $2t$. Hence, the average service latency for mobile device $m$ is given by
\begin{equation}\label{service_latency}
    \begin{split}
        \overline{D}_m (\theta_m) = \frac{\theta_m-1}{\theta_m}\cdot t+\frac{1}{\theta_m}\cdot(2t) = t\bigg(1 + \frac{1}{\theta_m}\bigg).
    \end{split}
\end{equation}


We now derive the QoD of mobile device $m$ based on (\ref{AoI}) and (\ref{service_latency}). Since large AoI or service latency both affect the sensing data quality, we define the impact functions of the AoI $G(\overline{A}_m)$ and the service latency $H(\overline{D}_m)$ as follows \cite{zhou2021towards}:
\begin{equation}
    \begin{split}
        G(\overline{A}_m) &= \overline{A}_{max}  - \overline{A}_m,\\
        H(\overline{D}_m) &= \overline{D}_{max} - \overline{D}_m,
    \end{split}
\end{equation}
where $\overline{A}_{max}$ and $\overline{D}_{max}$  represent the maximum permissible values for AoI and service latency, respectively. 


Thus, based on \cite{kang2023blockchain,zhou2021towards}, the QoD of mobile device $m$ is defined as
\begin{equation}
    QoD_m(\theta_m) = \ln\big(\alpha_m (G(\overline{A}_m) - H(\overline{D}_m)) + H(\overline{D}_m)+1\big),
\end{equation}
where $\alpha_m \in [0,1]$ is the preference value of mobile device $m$ for AoI over service latency. 


\subsection{Problem Formulation}

\subsubsection{Utility of mobile devices}
In the context of big data sharing for mobile AIGC services, the utility of mobile device $m$ is quantified as the disparity between its received reward $r_m$ and its cost incurred during data sharing\cite{wen2023task}. The cost for mobile device $m$ is defined as $(\xi_mf_m)$\cite{kang2023blockchain,wen2023freshness}, where $f_m = 1/\theta_m$ denotes the update frequency of mobile device $m$ and $\xi_m$ represents the cost of each update\cite{kang2023blockchain}. Hence, the utility of mobile device $m$ is given by
\begin{equation}
    U_m = r_m - \xi_mf_m.
\end{equation}
Because of information asymmetry, the BS lacks precise knowledge of the update costs associated with each mobile device\cite{wen2023freshness, wen2023task}, but it can divide $M$ mobile devices into discrete types by leveraging statistical distributions of mobile device types derived from historical data\cite{kang2023blockchain}, denoted as $\mathcal{K} = \small\{\phi_k:1\leq k \leq K\small\}$. We define the $k$-th type as $\phi_k \triangleq \frac{1}{\xi_k}$, where a smaller update cost corresponds to a higher mobile device type, and the mobile device types are organized as $\phi_1 \leq \phi_2 \leq \cdots \leq \phi_K$. For ease of presentation, we designate the mobile device with type $k$ as the type-$k$ mobile device. Thus, the utility of the type-$k$ mobile device is given by
\begin{equation}
    u_k (f_k, r_k) = r_k - \frac{f_k}{\phi_k}.
\end{equation}

\subsubsection{Utility of the BS}
Because of information asymmetry, the BS possesses knowledge solely regarding the total number of mobile devices and the type distribution but lacks information regarding the precise type of each individual mobile device\cite{wen2023task, kang2023blockchain}. Hence, the overall utility of the BS is given by\cite{kang2023blockchain,wen2023freshness,wen2023generative}
\begin{equation}\label{BS}
    U_B (\boldsymbol{f},\boldsymbol{r}) = M\sum_{k=1}^KQ_k(\beta QoD_k-r_k).
\end{equation}
Here, $\beta > 0$ represents the unit profit associated with QoD, $Q_k$ is the probability that a mobile device belongs to type-$k$, with the constraint that the sum of probabilities of all types is $1$. Additionally, $\boldsymbol{r} = [r_k]_{1\times K}$ and $\boldsymbol{f} = [f_k]_{1\times K}$ denote the vectors of rewards and update frequencies for all $K$ types of mobile devices, respectively.


\subsubsection{Contract formulation}
Since rational mobile devices may provide low-quality data to the BS to cheat higher rewards\cite{wen2023task}, we propose a contract theoretic model to mitigate the impact of information asymmetry\cite{zhou2021towards,kang2023blockchain}, thus motivating mobile devices to contribute high-quality sensing data for guaranteeing the AIGC service quality.

The BS designs a contract consisting of a group of contract items and provides them to mobile devices. The contract items are update frequency-reward pairs, denoted by $\Psi = \small\{(f_k, r_k),\:k\in \mathcal{K}\small\}$, where the reward $r_k$ paid to the type-$k$ mobile devices as an incentive according to its update frequency $f_k$\cite{kang2023blockchain}. To guarantee that each mobile device selects the contract item most advantageous for its specific type, the contract must satisfy both Individual Rationality (IR) and Incentive Compatibility (IC) constraints\cite{wen2023freshness,kang2023blockchain,zhou2021towards}:
\begin{definition}
    (Individual Rationality) The type-$k$ mobile devices can obtain a non-negative utility by choosing the contract item $(f_k, r_k)$, given by 
    \begin{equation}\label{IR}
    r_k - \frac{f_k}{\phi_k}\geq 0,\: \forall k \in \mathcal{K}.
\end{equation}
\end{definition}

\begin{definition}
(Incentive Compatibility) The type-$k$ mobile devices prefer to choose the contract item $(f_k, r_k)$ that is optimally tailored to their characteristics, given by
\begin{equation}\label{IC}
        r_k - \frac{f_k}{\phi_k} \geq r_j - \frac{f_j}{\phi_k},\:\forall k,j\in\mathcal{K},\: k \neq j.
\end{equation}
\end{definition}

With the IC and IR constraints, the BS endeavors to maximize its utility\cite{kang2023blockchain,zhou2021towards,wen2023freshness}, which is expressed as
\begin{equation}\label{problem1}
    \begin{split}
        &\max\limits_{\boldsymbol{f},\boldsymbol{r}}\:U_B (\boldsymbol{f},\boldsymbol{r}) \\
        &\:\:\text{s.t.}\:\: r_k - \frac{f_k}{\phi_k} \geq 0,\:\forall k \in \mathcal{K},\\
        &\qquad r_k - \frac{f_k}{\phi_k} \geq r_j - \frac{f_j}{\phi_k}, \:\forall k, j \in \mathcal{K},\\
        &\qquad f_k \geq 0, r_k \geq 0, \phi_k > 0,\: \forall k \in \mathcal{K}.
    \end{split}
\end{equation}
It is important to acknowledge that the aforementioned problem is intricate, entailing $N$ IR and $N(N-1)$ IC constraints. Besides, in the actual environment of mobile AIGC networks, the parameters in the constraints in (\ref{problem1}) change dynamically. Thus, it is challenging to effectively solve the optimization problem (\ref{problem1}) by using traditional mathematical techniques. Fortunately, DRL-based approaches can offer a feasible alternative\cite{zhang2023learning}, and we adopt the PPO algorithm that is robust and scalable to find the optimal contract, as elucidated in the subsequent section. 

\section{PPO-based Optimal Contract Design}
In this section, we first model the contract design between the BS and mobile devices as a Markov Decision Process (MDP). Then, we propose a PPO-based learning algorithm to find the optimal contract, where the BS is the learning agent.

\subsection{MDP Formulation}
\subsubsection{State space} In contract modeling, the state space that affects the optimal contract design at the current decision round $z\:(z = 1, 2, \ldots, Z)$ is defined as
\begin{equation}
\begin{split}
    \boldsymbol{s}^{(z)} \triangleq\{M, K, \overline{A}_{max}^{(z)}, \overline{D}_{max}^{(z)}, \mathcal{Q}^{(z)}, \mathcal{K}^{(z)}\},
\end{split}
\end{equation}
where $M$ and $K$ are constant, $\overline{A}_{max}^{(z)}$, $\overline{D}_{max}^{(z)}$, $\mathcal{Q}^{(z)} = (Q_1^{(z)},\ldots,Q_K^{(z)})$, and $\mathcal{K}^{(z)} = (\phi_1^{(z)},\ldots,\phi_K^{(z)})$ are generated randomly at the round $z$.

\subsubsection{Action space} Considering that the BS designs the contract items $\Psi$ to incentivize mobile devices to provide high-quality data, the action $\boldsymbol{a}^{(z)}$ at the round $z$ is defined as
\begin{equation}
\boldsymbol{a}^{(z)} \triangleq \{\Psi^{(z)}\},
\end{equation}
where $\Psi^{(z)} = \small\{(f_k^{(z)}, r_k^{(z)}),\:k\in \mathcal{K}\small\}$.

\subsubsection{Immediate reward} After taking the action $\boldsymbol{a}^{(z)}$, the BS can achieve the immediate reward $R (\boldsymbol{s}^{(z)}, \boldsymbol{a}^{(z)})$ to maximize its overall utility (\ref{BS}) while considering the IR (\ref{IR}) and IC (\ref{IC}) constraints. Thus, the reward function can be formulated as
\begin{equation}\label{Reward}
    R (\boldsymbol{s}^{(z)}, \boldsymbol{a}^{(z)})=\begin{cases}U_B^{(z)} (\boldsymbol{f},\boldsymbol{r}),&\text{if}\:\boldsymbol{a}^{(z)} \text{satisfies (\ref{IR}) and (\ref{IC})},
\\[2ex]U_{p},&\text{otherwise},\end{cases}
\end{equation}
where $U_B^{(z)} (\boldsymbol{f},\boldsymbol{r})$ is the overall utility of the BS at the round $z$ and $U_{p}$ is the penalty for not satisfying either IR or IC constraints. Note that $U_{p}$ is a negative value.

\subsection{PPO Algorithm for Optimal Contract Design}
PPO is a model-free and on-policy algorithm that iteratively updates the policy until optimal by using a clipped surrogate objective\cite{zhang2023energy}. Different from other DRL algorithms, PPO exhibits a more stable learning progression due to its conservative update strategy, which is especially useful for complex and high-dimensional problems. Given a \textit{contract design policy} $\pi_{\omega}$, which is parameterized by a Deep Neural Network (DNN) with $\omega$, the state-value function $V_{\pi_\omega}(\boldsymbol{s})$ measures the expected return from $\boldsymbol{s}^{(1)}$ to $\boldsymbol{s}^{(Z)}$\cite{zhang2023learning, zhang2023energy}, which is defined as 
\begin{equation}
    V_{\pi_\omega}(\boldsymbol{s}) = \hat{\mathbb{E}}_{\pi_{\omega}}\Bigg(\sum_{z=1}^{Z} \gamma^{z} R (\boldsymbol{s}^{(z)}, \boldsymbol{a}^{(z)}) \mid \boldsymbol{s}^{(1)}=\boldsymbol{s}\Bigg),
\end{equation}
where $\hat{\mathbb{E}}_{\pi_{\omega}}(\cdot)$ is the expected value of a random variable under the assumption that the agent adheres to policy $\pi_\omega$ and $\gamma \in [0,1]$ is the reward discount factor.

At each training iteration, the network parameters $\omega$ are updated by randomly sampling experiences from the replay buffer\cite{zhang2023learning,zhang2023energy}. Then, the variance-reduced advantage function estimator $A(\boldsymbol{s}, \boldsymbol{a})$ is calculated by using the \textit{generalized advantage estimation} based on the state-value function $V_{\pi_\omega}(\boldsymbol{s})$\cite{schulman2015high}, which is given by
\begin{equation}
\begin{split}
    A(\boldsymbol{s}^{(z)}, \boldsymbol{a}^{(z)}) = &\gamma^{Z-z}V_{\pi_{\omega_e}}(\boldsymbol{s}^{(Z)}) - V_{\pi_{\omega_e}}(\boldsymbol{s}^{(z)})\\
    &+\sum_{y = z}^{Z-1}\gamma^{y-z}R(\boldsymbol{s}^{(y)}, \boldsymbol{a}^{(y)}),
\end{split}
\end{equation}
where $\omega_e$ is the policy parameter in the episode $e$.

To train the actor and critic networks, the loss functions consist of the policy surrogate $L_C(\omega)$ and the value function error term $L_V(\omega)$\cite{zhang2023learning}. Specifically, $L_V(\omega)$ is given by
\begin{equation}
    L_V^{(z)}(\omega_e) = \Big(V_{\pi_{\omega_e}}(\boldsymbol{s}^{(z)})-V_{targ}^{(z)}\Big)^2,
\end{equation}
where $V_{targ}^{(z)}$ is the total discount reward from the round $z$ until the end of the episode. $L_C(\omega)$ is given by
\begin{equation}
\begin{split}
    L_C^{(z)}(\omega_e) = \hat{\mathbb{E}}\bigg(\min \Big(&r^{(z)}(\omega_e) A(\boldsymbol{s}^{(z)}, \boldsymbol{a}^{(z)}),\\ &g(\epsilon, r^{(z)}(\omega_e)) A(\boldsymbol{s}^{(z)}, \boldsymbol{a}^{(z)})\Big)\bigg),
\end{split}
\end{equation}
where $r^{(z)}(\omega_e)$ represents the important ratio between the new policy and the old policy\cite{zhang2023energy}, given by
\begin{equation}
    r^{(z)}(\omega_e) = \frac{\pi_{\omega_e}(\boldsymbol{a}^{(z)}\mid \boldsymbol{s}^{(z)})}{\pi_{\omega_e^{old}}(\boldsymbol{a}^{(z)}\mid \boldsymbol{s}^{(z)})},
\end{equation}
and $g(\epsilon, r^{(z)}(\omega_e))$ is given by
\begin{equation}
    g(\epsilon, r^{(z)}(\omega_e))=  \begin{cases}1-\epsilon,\:r^{(z)}(\omega_e)<1-\epsilon,\\
r^{(z)}(\omega_e),\:1-\epsilon \leq r^{(z)}(\omega_e) \leq 1+\epsilon,\\1+\epsilon,\:r^{(z)}(\omega_e)>1+\epsilon.\end{cases}  
\end{equation}
Here, $\omega_e^{old}$ is the old policy parameter for sampling in the episode $e$ and $\epsilon$ is a hyperparameter. To update the policy and value functions, we maximize the objective function through stochastic gradient ascent, given by\cite{zhang2023learning}
\begin{equation}\label{objective}
    \omega_{e+1}=\arg \max _{\omega_{e}} \frac{1}{\left|N\right|} \sum_{\left|N\right|} \hat{\mathbb{E}}\Big(L_{C}^{(z)}\left(\omega_{e}\right)-c L_{V}^{(z)}\left(\omega_{e}\right)\Big),
\end{equation}
where $\omega_{e+1}$ is the policy parameter in the episode $(e+1)$, $c$ is the loss coefficient, and $N$ is the batch size of experiences sampled for calculating the policy gradients.

Inspired by the preceding analysis, we present a PPO-based algorithm for optimal contract design, which is depicted in \textbf{Algorithm 1}. The computational complexity of the proposed algorithm is $\mathcal{O}\big(\sum_{p=1}^P n_{p-1}n_p\big)$\cite{zhang2023learning, zhang2023energy}, where $n_p$ represents the number of neural units at the $p$-th layer in the DNNs.

\begin{algorithm}[t]  
\small
\label{algorithm2}
\caption{PPO-based Optimal Contract Design}\label{algorithm}
Initialize number of rounds in an episode $Z$, number of episodes $E$, batch size $N$, network parameters $\omega$, reward discount factor $\gamma$, hyperparameter $\epsilon$\;
\For{Episode $e \in 1,\ldots,E $}
{   
    Reset environment state $\boldsymbol{s}^{(0)}$ and replay buffer $\mathcal{D}$\;
    \For{Round $z \in 1,\ldots,Z$}
    {   
        Observe the current environment $\boldsymbol{s}^{(z)}$;\\
        Take action $\boldsymbol{a}^{(z)}$ based on the policy $\pi_{\omega_e}$;\\
        Update $\boldsymbol{s}^{(z)}$ into $\boldsymbol{s}^{(z+1)}$ and calculate reward $R^{(z)}$ through (\ref{Reward});\\
        Store transition $(\boldsymbol{s}^{(z)},\boldsymbol{a}^{(z)},R^{(z)},\boldsymbol{s}^{(z+1)})$ into
 $\mathcal{D}$\;
        \If{ $z\%\left|N\right|==0$}
        {
            \For{$x \in 1,\ldots,X $}
            {
                Sample a random mini-batch of data with a size $\left|N\right|$ from $\mathcal{D}$ to update the actor and critic networks through (\ref{objective});
            }
        }
    }
}
\end{algorithm}

\begin{figure}[t]
\centering
\captionsetup{font=footnotesize}
\subfigure[Test rewards of the proposed PPO-based algorithm with different seeds and the random algorithm under information asymmetry, with batch size $N = 512$, discount factor $\gamma = 0.95$, and the learning rate of actor and critic networks is $10^{-4}$.]{
\centering
\includegraphics[width=0.4\textwidth]{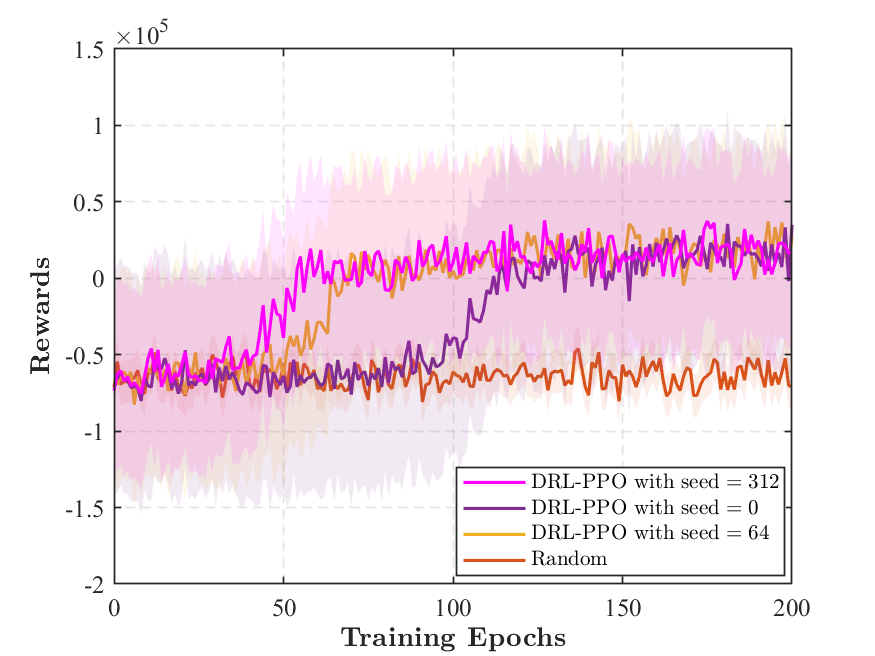}
\label{Compare}
}
\captionsetup{font=footnotesize}
\subfigure[Optiaml contract designs under different network states.]{
\centering
\includegraphics[width=0.4\textwidth]{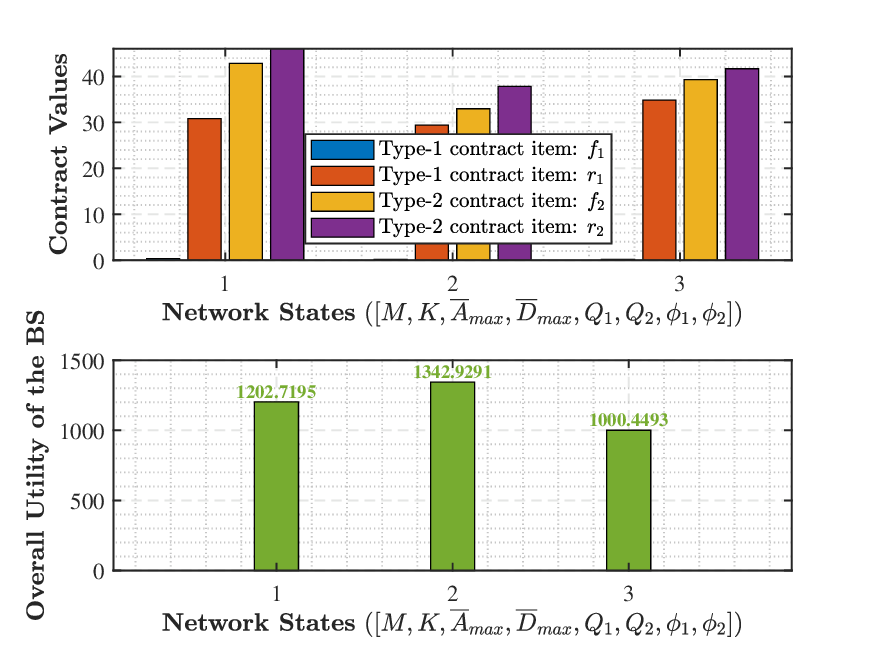}
\label{Contract_design}
}
\captionsetup{font=footnotesize}
\caption{Performance analysis of the proposed PPO-based algorithm for optimal contract design, where $\alpha = 0.75$ and the seed is set to $312$.}
\label{PPO_performance}
\end{figure}

\begin{figure}[t]
\centerline{\includegraphics[width=0.4\textwidth]{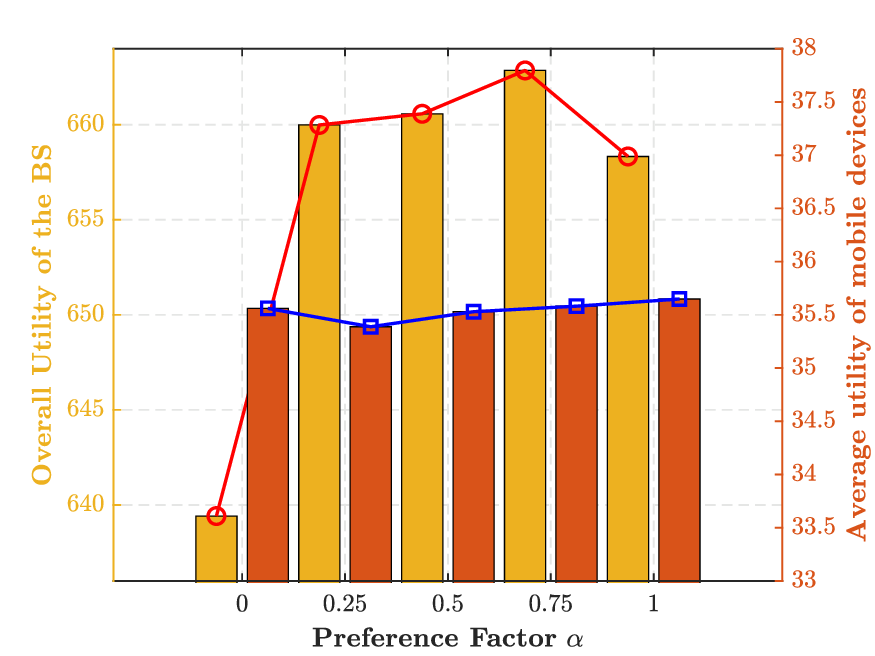}}
\captionsetup{font=footnotesize}
\caption{Utilities of the BS and mobile devices under different $\alpha$.}
\label{Utility}
\end{figure}

\section{Numerical Results}
In this section, we first evaluate the performance of the proposed PPO-based algorithm for optimal contract design. Then, we analyze the effect of preference factor $\alpha$ on utilities of the BS and mobile devices. We consider $40$ mobile devices with identical characteristics, which are divided into $2$ types. According to \cite{zhang2018towards, kang2023blockchain, wen2023generative, wen2023freshness, zhou2021towards}, $\phi_1$ and $\phi_2$ are randomly sampled within $[1,8]$ and $[8,15]$ based on the magnitude of update cost, respectively. $Q_1$ and $Q_2$ are randomly generated following the Dirichlet distribution. The preference factors $\alpha$ of all mobile devices are identical. The maximum tolerant AoI $\overline{A}_{max}$ and service latency $\overline{D}_{max}$ are both randomly sampled within $[0.5,1]$. The unit profit $\beta$ is set to $100$. The penalty for not satisfying either IR or IC constraints $U_p$ is set to $-500$. Without loss of generality, we set $l = 24\:\rm{Kb}$ and $\tau = 24\:\rm{Mbps}$, resulting in $t = 1\:\rm{ms}$\cite{zhang2018towards}. Note that the experiments are carried out using PyTorch with CUDA 12.0 on NVIDIA GeForce RTX 3080 Laptop GPU.

Figure \ref{PPO_performance} presents the performance analysis of the proposed PPO-based algorithm for optimal contract design under information asymmetry. We compare the proposed PPO-based algorithm with the $\textit{Random algorithm}$ that the BS randomly designs contracts regardless of the types of mobile devices. As shown in Fig. \ref{Compare}, the proposed PPO-based algorithm outperforms the random algorithm, indicating that the proposed PPO-based can more effectively find the optimal contract. The reason is that PPO typically requires fewer samples, which can achieve good performance with a relatively small number of interactions with the environment. This characteristic is crucial for mobile AIGC networks where data collection may be expensive or time-consuming\cite{lai2023resource}. Figure \ref{Contract_design} illustrates the details of the designed type-$1$ and type-$2$ contracts under three network states, where state 1 is $[40, 2, 0.95, 0.73, 0.84, 0.16, 2, 12]$, state 2 is $[40, 2, 0.94, 0.85, 0.80, 0.20, 2, 12]$, and state 3 is $[40, 2, 0.95, 0.81, 0.43, 0.57, 2, 13]$. We can observe that the proposed PPO-based algorithm maintains stable performance regardless of network states, which can adapt to the complexity of mobile AIGC networks\cite{lai2023resource}.


Figure \ref{Utility} depicts the effects of the preference factor $\alpha$ on the utilities of the BS and mobile devices. We can observe that as $\alpha$ increases, the overall utility of the BS initially rises before subsequently declining. Thus, there exists an optimal value of $\alpha$ that maximizes the utility of the BS. Besides, as $\alpha$ increases, the average utility of mobile devices is stable compared with the overall utility of the BS. This phenomenon arises because the elevation of $\alpha$ indicates that mobile devices prioritize the reduction of AoI in data sharing rather than service latency\cite{kang2023blockchain}, leading to an increase in the update frequency of mobile devices, thereby ensuring the freshness of sensing data. Under the action of contracts, the increase in update frequency enables mobile devices to receive more rewards, which results in the stability of the average utility of mobile devices. In summary, the above results demonstrate the superior performance of the proposed PPO-based algorithm for optimal contract design. 

\section{Conclusion}
In this paper, we investigated the provision of AIGC services for mobile devices and devised an incentive mechanism admist information asymmetry in mobile AIGC networks. Specifically, we first developed a quality of data metric by jointly considering service latency and AoI in big data sharing, where the AoI is a novel metric capable of effectively quantifying data freshness. To incentivize mobile devices to provide high-quality sensing data for AIGC model fine-tuning, we proposed a contract theoretic model. Furthermore, we adopted the PPO algorithm for optimal contract design. Finally, numerical results demonstrate the reliability and effectiveness of the proposed PPO-based contract model. For future work, we will focus on designing reliable incentive mechanisms between multiple mobile devices and multiple BSs for mobile AIGC service provisions. Furthermore, we will employ advanced AI tools like Generative Diffusion Models (GDMs) and Mixture of Experts (MoE) to enhance the solution methodology.

\bibliographystyle{IEEEtran}
\bibliography{ref}
\end{document}